\documentclass[fleqn,twoside]{article}
\usepackage{espcrc2}
\usepackage{graphicx}
\usepackage[figuresright]{rotating}

\newcommand{\text}[1]{\textrm{#1}}

\newcommand{\AmS}{{\protect\the\textfont2
  A\kern-.1667em\lower.5ex\hbox{M}\kern-.125emS}}
\begin{document}

\title{The spin-dependent nd scattering length - a proposed high-accuracy
  measurement\thanks{nucl-ex/0401029. To be published in
    \emph{Nucl.~Inst.~Methods~}\textbf{A} as part of the Proceedings of the
    \emph{9th International Workshop on Polarized Solid Targets and
      Techniques} in Bad Honnef (Germany), 27th -- 29th October 2003.}  }

\author{B. van den Brandt%
  \address{Paul Scherrer Institute, CH-5232
        Villigen PSI, Switzerland}, H. Gl\"{a}ttli%
      \address{Commissariat \`{a} l'Energie Atomique, CE Saclay/SPEC LLB,
        F-91191 Gif-sur-Yvette, France},
      H.W. Grie\ss hammer%
      \address{Physik-Department, Technische Universit\"{a}t M\"{u}nchen,
        D-85748 Garching, Germany},
      P. Hautle${}^a$, J. Kohlbrecher${}^a$, J.A. Konter${}^a$, \\
and O. Zimmer${}^c$\thanks{%
  Corresponding author, oliver.zimmer@ph.tum.de}}

\begin{abstract}
  The understanding of few-nucleon systems at low energies is essential, e.g.
  for accurate predictions of element abundances in big-bang and stellar
  fusion. Novel effective field theories, taking only nucleons, or nucleons
  and pions as explicit degrees of freedom, provide a systematic approach,
  permitting an estimate of theoretical uncertainties. Basic constants
  parameterising the short range physics are derived from only a handful of
  experimental values. The doublet neutron scattering length $a_{2}$ of the
  deuteron is particularly sensitive to a three-nucleon contact interaction,
  but experimentally known with only $6$ \% accuracy. It can be deduced from
  the two experimentally accessible parameters of the nd scattering length. We
plan to measure the poorly known "incoherent" nd scattering length $a_{%
  \mathrm{i,d}}$ with $10^{-3}$ accuracy, using a Ramsey apparatus for
pseudomagnetic precession with a cold polarised neutron beam at PSI. A
polarised target containing both deuterons and protons will permit a
measurement relative to the incoherent np scattering length, which is know
experimentally with an accuracy of $2.4\times 10^{-4}$.\smallskip

 \vspace{1pc}
\end{abstract}

\maketitle


\section{Introduction}

In the past few years, a new strategy has been developed to describe nuclear
forces at low energy. Chiral perturbation theory ($\chi $PT) is an effective
field theory, which describes interactions of pions and between pions and
nucleons (N). It leads to a systematic expansion of the scattering amplitude
in powers of ratios of small momenta and low-energy input parameters like the
pion mass over the breakdown scale of the theory. For the first time the
accuracy of calculations can be estimated in a model-independent theory of
nuclear interactions, providing reliable predictions of many important
low-energy quantities. These are ground state properties of bound systems and
processes involving external and exchange currents, as e.g. cross
sections relevant for big-bang nucleosynthesis and stellar fusion \cite%
{Nollet/2000,Burles/1999}. Also the determination of fundamental properties of
the neutron from experiments on few-nucleon systems mandates a
model-independent subtraction of nucleon binding and meson exchange effects.

Weinberg pointed out that chiral three-nucleon (3N) forces appear naturally in
$\chi $PT \cite{Weinberg/1991}. The most relevant processes are: a two-pion
exchange, a 2N contact interaction with pion exchange, and a 3N contact
interaction \cite{Kolck/1994}. The contact interactions parameterise the
short-range physics. As in Fermi's theory of weak interaction, they are
characterised by effective couplings, called low-energy constants (LECs).
They have to be fixed by measured data of two independent low-energy 3N
observables. Very recently, a first complete analysis of nd scattering at
next-to-next-to leading order has been performed with impressive results
\cite{Epelbaum/2003}. All observables are expanded in powers of momenta and
the pion mass over the $\chi $PT-breakdown scale of about $800$ MeV.

An even simpler approach to the nuclear few-body problem is an effective
field theory without pionic degrees of freedom \cite{Chen/1999,Bedaque/2002}%
. This theory starts out from point-like interactions between nucleons, which
only have to respect the symmetries of QCD. Like $\chi $PT, it describes
phenomena in a systematic way, but is applicable only at energies well below a
breakdown scale set by the pion mass. Again, only two LECs characterising 3N
forces are required to predict observables with an accuracy of less than $1$
\% in processes involving three nucleons.

\begin{figure}
\begin{center}
  \includegraphics*[width=\linewidth]{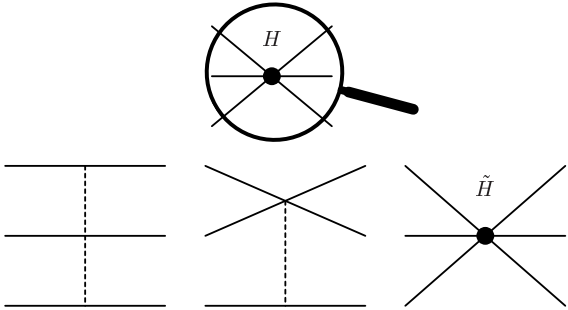}
 \caption{The strengthes $H$ of the point-like
   three-body forces of the effective field theory in which pions are
   integrated out (top) must be determined from experimental three-body data.
   At momenta above $m_{\protect\pi }c$, they are partially resolved as
   pion-exchange with known couplings (bottom), but the core-strengthes
   $\widetilde{H}$ still need the input from three-nucleon observables. Solid
   (dashed) lines denote nucleons (pions).}
\end{center}
\end{figure}

However, this accuracy can only be achieved if the experimental inputs are
known correspondingly well. The binding energy of the triton and the doublet
nd scattering length $a_{2}$ are particularly well suited to determine the
LECs for $\chi $PT and for the pion-free theory. First, there are no Coulomb
effects to be considered. Second, $a_{2}$ is very sensitive to 3N forces: in
the quartet channel, the incident neutron has its spins parallel to the one
bound in the deuteron, so that the Pauli principle prohibits 3N forces to
play any sizeable role at low momentum transfer. In contradistinction, the $%
s $-wave in the doublet channel allows for a momentum-independent 3N
interaction. This turns out even necessary to achieve results which are not
sensitive to physics at high energy scales, or respectively, at short
distances beyond the range of applicability of the theory. While the triton
binding energy is known with an accuracy of $5\times 10^{-7}$, the
experimental knowledge of the nd doublet scattering length is only $6$~\%.

\section{Present situation and accuracy goal}

The scattering length of a neutron with spin $\mathbf{s}$ and a nucleus with
spin $\mathbf{I}$ is given by%
\[
a=\frac{I+1}{2I+1}a_{+}+\frac{I}{2I+1}a_{-}+\frac{2\left( a_{+}-a_{-}\right)
}{2I+1}\,\mathbf{s\cdot I\medskip }
\]%
\begin{equation}
\quad =a_{\mathrm{c}}+\frac{2a_{\mathrm{i}}}{\sqrt{I\left( I+1\right) }}\,%
\mathbf{s\cdot I,}
\end{equation}%
where $a_{+}$ and $a_{-}$ denote the scattering lengths in the state with
total spin $I+\frac{1}{2}$, respectively, $I-\frac{1}{2}$. Since one cannot
prepare the latter state, it is not possible to measure $a_{-}$ directly.
Experimentally accessible are the spin-independent, coherent scattering length
$a_{\mathrm{c}}$, and the factor $a_{\mathrm{i}}$ which parameterises the
spin-dependence (sometimes called "incoherent" scattering length). For
the neutron-deuteron system, the doublet and quartet scattering lengths, $%
a_{-}=a_{2}$, respectively $a_{+}=a_{4}$, are given as the linear combinations
\[
a_{2}=a_{\mathrm{c,d}}-\sqrt{2}\,a_{\mathrm{i,d}},
\]%
\begin{equation}
a_{4}=a_{\mathrm{c,d}}+\frac{1}{\sqrt{2}}\,a_{\mathrm{i,d}}.  \label{a2a4}
\end{equation}%
The best experimental value of $a_{2}$ was obtained 30 years ago \cite%
{Dilg/1971}, using a combination of a measurement of the scattering cross
section of the free deuteron
\begin{equation}
\sigma _{\mathrm{s,d}}=4\pi \left( \left\vert a_{\mathrm{c,d}}\right\vert
^{2}+\left\vert a_{\mathrm{i,d}}\right\vert ^{2}\right) ,
\end{equation}%
with a gravity refractometric measurement of the bound coherent nd scattering
length $b_{\mathrm{c,d}}$ (through the relation of bound and free
scattering lengths $b$, respectively $a$ by%
\begin{equation}
a=\frac{M}{M+m}b,  \label{a-b}
\end{equation}%
with corresponding indices. $M$ is the mass of the nucleus and $m$ the neutron
mass). The experimental values were $\sigma _{\mathrm{s,d}}=3.390\pm
0.012~\mathrm{barn}$ and $b_{\mathrm{c,d}}=6.672\pm 0.007~\mathrm{fm}$,
leading to%
\begin{equation}
a_{2}=0.65\pm 0.04~\mathrm{fm.}
\end{equation}

Recently, a new measurement of $b_{\mathrm{c,d}}$ was performed at the NIST
interferometer at Washington DC \cite{Schoen/2003}. Including the result, $%
b_{\mathrm{c,d}}=6.6649\pm 0.0040~\mathrm{fm}$, the present world average is
\begin{equation}
b_{\mathrm{c,d}}=6.6683\pm 0.0030~\mathrm{fm.}  \label{coherent average}
\end{equation}%
However, this improvement does not significantly reduce the experimental
uncertainty of $a_{2}$, since this is dominated by the unsufficient
knowledge of $a_{\mathrm{i,d}}$. On the other hand, the authors of ref.\cite%
{Schoen/2003} argue that, because $a_{4}$ should have a very small dependence
on 3N forces, $a_{\mathrm{i,d}}$ may be derived from the experimental value
stated in eq.(\ref{coherent average}) and a theoretical value of $a_{4}$,
using eq.(\ref{a2a4}) and eq.(\ref{a-b}). That way, they
obtain a semi-experimental value $a_{2}=0.645\pm 0.003\left( \textrm{expt}%
\right) \pm 0.007\left( \textrm{theory}\right) ~\mathrm{fm}$.

The goal of the present experiment is a direct measurement of $a_{\mathrm{i,d%
  }}$, which does not rely on any nuclear few-body theoretical input. As a
minimum aim we hope to achieve an accuracy of $10^{-3}$. Using eq.(\ref{a2a4}%
) together with eq.(\ref{coherent average}), this shall provide a new value of
$a_{2}$ with an uncertainty of $0.004~\mathrm{fm}$.

\section{Method}

The spin-dependent scattering length induces a spin-dependence of the neutron
refractive index. As a result, $b_{\mathrm{i}}$ can be determined directly
with a polarised neutron beam passing through a polarised target, via
detection of pseudomagnetic neutron precession around the axis of nuclear
polarisation \cite{Barychevsky/1965,Abragam/1973}. The pseudomagnetic
precession angle is given by
\begin{equation}
\varphi ^{\ast }=2\lambda d\sum\limits_{k}\sqrt{\frac{I_{k}}{I_{k}+1}}%
P_{k}N_{k}b_{\mathrm{i},k}.  \label{pseudoangle}
\end{equation}%
$N_{k}$ is the number density, $I_{k}$ the nuclear spin and $P_{k}$ the
nuclear polarisation, with the sum index $k$ extending over the nuclear
species with spin. $\lambda $ is the de-Broglie wavelength of the neutrons,
and $d$ is the thickness of the sample. The angle $\varphi ^{\ast }$ can be
measured with an accuracy of at least one degree, using the method described
in \cite{Glattli/1987,Glattli2/1987,Glattli3/1987} based on Ramsey's
well-known resonance technique with two separated oscillatory fields. The
target is situated in the homogeneous magnetic field between the two
high-frequency $\frac{\pi }{2}$ coils.

The uncertainty of the deuteron polarisation $P_{\mathrm{d}}$ would impose a
severe limitation in accuracy if the deuterons were the only nuclei with spin
in the target. This difficulty can be considerably relaxed in a measurement of
$b_{\mathrm{i,d}}$ relative to $b_{\mathrm{i,p}}$ of the proton, which is
known with the high accuracy of $2.4\times 10^{-4}$. Using a single target
which contains both deuterons and protons at hydrogen sites, absolute
polarisation measurements can be avoided \cite{Glattli/1983}. Many materials
are suitable to apply the method of dynamic nuclear polarisation (DNP), by
which both isotopes can be polarised simultaneously under still rather
convenient conditions \cite{Abragam/1982}.

The method combines several measurements described in the following. First,
the sample is polarised via DNP. After freezing the nuclear polarisation one
determines the pseudomagnetic precession angle. According to eq.(\ref%
{pseudoangle}) and including an additional, instrumental phase $\varphi _{0}$%
, it is given by
\begin{equation}
\phi _{1}=\varphi _{\mathrm{d}}^{\ast }+\varphi _{\mathrm{p}}^{\ast
}+\varphi _{0}
\end{equation}%
with%
\[
\varphi _{\mathrm{d}}^{\ast }=\sqrt{2}\lambda dP_{\mathrm{d}}N_{\mathrm{d}%
}b_{\mathrm{i,d}},
\]%
\begin{equation}
\varphi _{\mathrm{p}}^{\ast }=\frac{2}{\sqrt{3}}\lambda dP_{\mathrm{p}}N_{%
\mathrm{p}}b_{\mathrm{i,p}}.  \label{pseudoangles-d-p}
\end{equation}%
Further, using hf-saturation, one can selectively depolarise either the
protons or the deuterons, without significantly affecting the polarisation of
the other spin species. The subsequent cross relaxation between the two spin
systems can be held sufficiently slow by keeping the temperature
sufficiently low. Saturating, e.g. first the protons, one can then measure%
\begin{equation}
\phi _{2}=\varphi _{\mathrm{d}}^{\ast }+\varphi _{0}.
\end{equation}%
After a subsequent depolarisation of the deuterons one measures%
\begin{equation}
\phi _{3}=\varphi _{0}.
\end{equation}%
Combining the measured values and using eq.(\ref{pseudoangles-d-p}), we
obtain%
\begin{equation}
b_{\mathrm{i,d}}=\sqrt{\frac{2}{3}}\frac{\phi _{2}-\phi _{3}}{\phi _{1}-\phi
_{2}}\frac{P_{\mathrm{p}}N_{\mathrm{p}}}{P_{\mathrm{d}}N_{\mathrm{d}}}b_{%
\mathrm{i,p}}.  \label{rat}
\end{equation}

The method is completed with measurements of deuteron and proton NMR signals
$\mathcal{I}_{k}$, taken as integral of the corresponding rf-absorption line
before and after each measurement of a pseudomagnetic precession angle:%
\begin{equation}
\mathcal{I}_{k}=C_{k}P_{k}N_{k}.  \label{NMR}
\end{equation}%
Apart from natural constants, $C_{k}$ is given as the product of $%
g_{k}^{2}/I_{k}$ (with $g_{k}$ denoting the $g$-factor) \cite{Goldman/1975}
and a factor accounting for the sensitivity of the resonance circuit. The
latter would be difficult to determine absolutely. However, since only the
ratio of $P_{k}N_{k}$ for the two spin species occurs in eq.(\ref{rat}),
replacing this by the ratio of NMR signals via eq.(\ref{NMR}), the common
instrumental factor cancels out. This requires using the same, linear
resonance circuit at the same frequency in the measurements of $\mathcal{I}_{%
  \mathrm{d}}$ and $\mathcal{I}_{\mathrm{p}}$, which therefore have to be
performed at different main magnetic fields to account for the different
gyromagnetic ratios.

Combining the various measurements into ratios, we thus finally obtain%
\begin{equation}
b_{\mathrm{i,d}}=\frac{1}{\sqrt{6}}\frac{g_{\mathrm{d}}^{2}}{g_{\mathrm{p}%
}^{2}}\frac{\phi _{2}-\phi _{3}}{\phi _{1}-\phi _{2}}\frac{\mathcal{I}_{%
\mathrm{p}}}{\mathcal{I}_{\mathrm{d}}}b_{\mathrm{i,p}}.  \label{final}
\end{equation}%
It is the gist of this method that we donot suppose any exact knowledge of
neither $N_{\mathrm{d}}$, $N_{\mathrm{p}}$, $P_{\mathrm{d}}$, $P_{\mathrm{p}%
} $, $d$ or $\lambda $, nor of any absolute calibration factors.

\section{Some practical comments}

The choice of the sample is governed by several factors. First, we consider
the isotopic composition. Best sensitivity is attained for $\varphi _{%
  \mathrm{d}}^{\ast }\approx \varphi _{\mathrm{p}}^{\ast }$. DNP keeps the
spin temperatures of protons and deuterons equal \cite{de Boer/1974}. Using
the Brillouin functions in the high-temperature limit,%
\begin{equation}
\frac{P_{\mathrm{d}}}{P_{\mathrm{p}}}\approx \frac{4\gamma _{\mathrm{d}}}{%
3\gamma _{\mathrm{p}}}\approx 0.2.
\end{equation}%
From eq.(\ref{pseudoangles-d-p}),%
\begin{equation}
\frac{N_{\mathrm{d}}}{N_{\mathrm{p}}}\approx 5\,\frac{P_{\mathrm{p}}}{P_{%
\mathrm{d}}}\frac{\varphi _{\mathrm{d}}^{\ast }}{\varphi _{\mathrm{p}}^{\ast
}}.
\end{equation}%
Thus, $\varphi _{\mathrm{d}}^{\ast }\approx \varphi _{\mathrm{p}}^{\ast }$ for
$N_{\mathrm{d}}/N_{\mathrm{p}}\approx 4$ \%. On the other hand, to keep the
systematic uncertainties induced by non-linearities of the NMR resonance
circuit small, requires $\mathcal{I}_{\mathrm{d}}\approx \mathcal{I}_{%
  \mathrm{p}}$, which happens to be true for about the same ratio. Considering
a typical material with number density $6.7\times 10^{22}$ cm$^{-3}$ of
hydrogen sites and the measurements being done at $\lambda =0.4$ nm,\ the
corresponding pseudomagnetic precession angles are%
\begin{equation}
\varphi _{\mathrm{d}}^{\ast }=90~\text{rad\thinspace }P_{\mathrm{d}},\quad
\varphi _{\mathrm{p}}^{\ast }=440~\text{rad\thinspace }P_{\mathrm{p}}.
\end{equation}%
Since the Ramsey technique is sensitive to precession angles of at least one
degree, a nuclear polarisation of a few percent is sufficient already for a
target only $3$ mm long.

Relaxation times are known to be strongly temperature-dependent. Using a
magnetic field of $2.5$ T and a cryostat providing $T\approx 100$ mK, the
nuclear spin systems are frozen. Typical spin-lattice relaxation times of $%
500$ hours and cross relaxation time of $100$ hours, obtained for (CH$_{2}$%
OH)$_{2}$ doped with Cr(V) paramagnetic centres (radicals), were published in
ref. \cite{de Boer/1974}. These are sufficiently long, since, for the given
neutron intensity, a single run of the experiment will take much less than one
hour. Operating with an amount of radicals below the optimum value to reach
maximum polarisation (which is not required here), spin relaxation may be
suppressed even further.

Apart from stability requirements, keeping low the amount of radicals is also
of interest to suppress systematic errors which may be due to their local
magnetic fields. Nuclear spins close to the paramagnetic electrons contribute
to pseudomagnetic precession but usually stay undetected by NMR, since their
resonance frequencies are strongly shifted with respect to the NMR line of the
nuclei in the bulk. Two measures may be taken. First, the radical
concentration should be kept as low as possible. Second, deuterated radicals
should be used, in order to keep the few protons of the sample away from the
radicals. Also for practical reasons, a good candidate of a target material is
a plastic deuterated to the required abundance and doped with deuterated
nitroxyl radicals \cite{van den Brandt/2003}. Note in addition that, using a
novel stroboscopic technique of simultaneous small angle neutron scattering
and NMR measurements, detailed information can be obtained about the
polarisation state of protons close to radicals dissolved in a deuterated
matrix \cite{van den Brandt/2002}. That way one can determine a correction, if
necessary. Note also, that the paramagnetic electrons are always polarised
very close to $100$ \%, and their fields cause a small magnetic neutron
precession, which is included in the instrumental phase $\varphi _{0}$.

The NMR detection scheme is a crucial part of the setup and deserves special
consideration. An analysis of several possibilities is described in a separate
publication \cite{Hautle/2003}. Apart from its special role in the
measurement, NMR can also be used to monitor the polarisation state of the
spin systems at any stage of the experiment. Many systematic variations of
parameters can be performed to check the independence of the measurement with
respect to the conditions which have dropped out in the derivation of
eq.(\ref{final}). Different samples with different chemical composition,
varied degree of deuteration and polarisation will be used. A systematic
variation of target and beam parameters should be able to demonstrate the
reliability of the measurement (this strategy was also adopted in a recent
determination of the spin-dependent n$^{3}$He scattering length \cite%
{Zimmer/2002}).

\end{document}